\documentclass{PoS}

\leftline{\parbox{3cm}{\large\rm  DESY 06-177 \\HU-EP-06/33 \\ LMU-ASC 65/06}
\vspace{1cm}}

\title{The QCD vacuum probed by overlap fermions}
\ShortTitle{The QCD vacuum probed by overlap fermions}

\author{\speaker{Volker Weinberg}\\
  Deutsches Elektronen-Synchrotron DESY, 15738 Zeuthen, Germany\\
  Institut f\"ur theoretische Physik, Freie Universit\"at Berlin, 14196 Berlin, 
  Germany\\
  E-mail: \email{volker.weinberg@desy.de}}

\author{Ernst-Michael Ilgenfritz\\
  Institut f\"ur Physik, Humboldt Universit\"at zu Berlin, 12489 Berlin, Germany\\
  E-mail: \email{ilgenfri@physik.hu-berlin.de}}

\author{Karl Koller\\
  Sektion Physik, Universit\"at M\"unchen, 80333 M\"unchen, Germany\\
  E-mail: \email{karl.koller@lrz.uni-muenchen.de}}

\author{Yoshiaki Koma\\
  Deutsches Elektronen-Synchrotron DESY, 22603 Hamburg, Germany\\
  E-mail: \email{yoshiaki.koma@desy.de}}

\author{Gerrit Schierholz\\
 Deutsches Elektronen-Synchrotron DESY, 22603 Hamburg, Germany\\
 John von Neumann-Institut f\"ur Computing NIC, 15738 Zeuthen, Germany \\
 E-mail: \email{gerrit.schierholz@desy.de}}

\author{Thomas Streuer\\
  Department of Physics and Astronomy, University of Kentucky, Lexington, 
  KY 40506-0055, USA \\ 
  E-mail: \email{streuer@pa.uky.edu}}

\abstract{
We summarize different uses of the eigenmodes of the Neuberger
overlap operator for the \linebreak analysis of the QCD vacuum, here applied to quenched
configurations simulated by means of the L\"uscher-Weisz action.
We describe the localization and chiral properties of the lowest modes. 
The overlap-based topological charge density (with and without 
UV-filtering) is compared with the results of UV-filtering for 
the field strength tensor. The latter allows to identify domains of 
good \linebreak (anti-)selfduality. All these techniques together lead to a dual 
picture of the vacuum, unifying the infrared instanton picture with the 
presence of singular defects co-existent at different scales.
}

\FullConference{XXIVth International Symposium on Lattice Field Theory\\
                July 23-28, 2006\\
                Tucson, Arizona, USA}

\newcommand{\Tr}{\mbox{Tr}}

\begin{document}

\section{Introduction}

The understanding of the vacuum structure of QCD is an important part of 
theoretical particle physics. The distribution of the topological charge density 
is believed to describe nonperturbative phenomena like the large $\eta'$ mass
through the axial U(1) anomaly
and the spontaneous 
breaking of chiral symmetry. The relation to confinement is less obvious.
The instanton picture models the QCD vacuum as a random (quenched)
or interacting (full QCD) instanton-antiinstanton system, whereas the
understanding of confinement seems to require singular fields emanating 
from lower-dimensional defects~\cite{Zakharov}. 

Recent advances in implementing chiral symmetry on the lattice have made 
it possible to find, from  first principles, manifestations of these 
dual aspects of vacuum structure. Overlap fermions are described by the 
Neuberger-Dirac operator $D_N$
constructed as a solution of the 
Ginsparg-Wilson relation. They are the cleanest known implementation  
of lattice fermions. For gauge fields free of dislocations, the index theorem 
is unambiguously realized
and the spectrum consists of chiral zero modes and 
non-chiral non-zero modes. 
While the number and chirality of zero modes provides a value for the total topological 
charge, the non-chiral non-zero modes appear in pairs of eigenvalues 
$\pm i \lambda$.

For our methodical study we use quenched configurations simulated by means of 
the L\"uscher-Weisz action. In the following table we list the statistics of 
lattices used in our investigation. We show also the topological susceptibility
$\chi_t$.
To set the scale, we use the quenched 
pion decay constant $f_{\pi}=91(2)$ MeV.

\begin{table}[h]
%\vspace*{0.1cm}
\begin{center}
\begin{tabular}{|c|c|c|c|r|r|r|c|}
\hline
$\beta$ & $a$ [fm] & lattice size    & $L$ [fm] & $V$ [fm$^4$] & $\chi_t$ [fm$^{-4}$] & confs. & modes    \\
\hline
\hline
8.45    & 0.105(2) & $12^3\times 24$ & 1.26 &  5.04 & 0.517(38) &  437 & 50 \\
8.45    & 0.105(2) & $16^3\times 32$ & 1.68 & 15.93 & 0.533(38) &  400 & 136-151\\
8.45    & 0.105(2) & $24^3\times 48$ & 2.52 & 80.66 & 0.522(48) &  250 & 144-177\\
\hline
8.10    & 0.138(3) & $12^3\times 24$ & 1.66 & 15.04 & 0.647(58) &  251 & 137-154\\
\hline
8.00    & 0.154(4) & $16^3\times 32$ & 2.46 & 73.72 & 0.607(21) & 2156 & 160-178\\
\hline
\end{tabular}
\end{center}
\label{tab:ensembles}
%\caption{Simulation parameters}
\end{table}

After presenting results concerning localization and chiral properties of
individual eigenmodes in different parts of the spectrum 
up to $\lambda = 400$ MeV, we describe two specific applications of 
the overlap Dirac operator. One is aiming to extract a UV-filtered 
field strength tensor~\cite{Gattringer} in order to assess the local degree 
of (anti-)selfduality. The other uses the fermionic definition of topological 
charge density~\cite{Laliena}. 
%% We distinguish between the all-scale density 
%% $q(x)$ containing fluctuations of all scales and the mode-truncated version 
%% thereof, $q_{\lambda_{\rm cut}}(x)$.
We interpret the respective structures and their dimensionality. 
%We find that 
%the coarse-grained features seem to be in accordance with an effective picture
%of nearly (anti-)selfdual lumps, whereas the all-scale density shows nested 
%structures with all fractal localization dimensions $d^{*} \leq 3$
%embedded in the global three-dimensional sign-coherent structure discovered by 
%Horvath {\it et al.}~\cite{Horvath}.

\section{Localization, dimensionality and local chirality of the lowest 
eigenmodes}

Recently, the localization, the eventually lower dimensionality 
and the local chirality of the lowest eigenmodes of the Dirac operator 
have attracted a lot of interest~\cite{MILC,Gubarev}. 
The reason for this activity is that these modes might be pinned down 
by some topological defects. This is inspired by certain confinement 
mechanisms, e.g. the center vortex mechanism. 
In model field configurations of thick vortices the zero modes have been 
seen~\cite{Gattnar} localized along the vortices with additional peaks at 
the intersections. The particular interest in the lowest eigenmodes is 
motivated by the observation that the lightest hadron propagators are very 
well approximated by taking only $O(50)$ lowest eigenmodes in the quark 
propagator into account.

\begin{figure}[ht]
\begin{center}
\begin{tabular}{cc}
\hspace*{-1cm}\includegraphics[width=7cm]{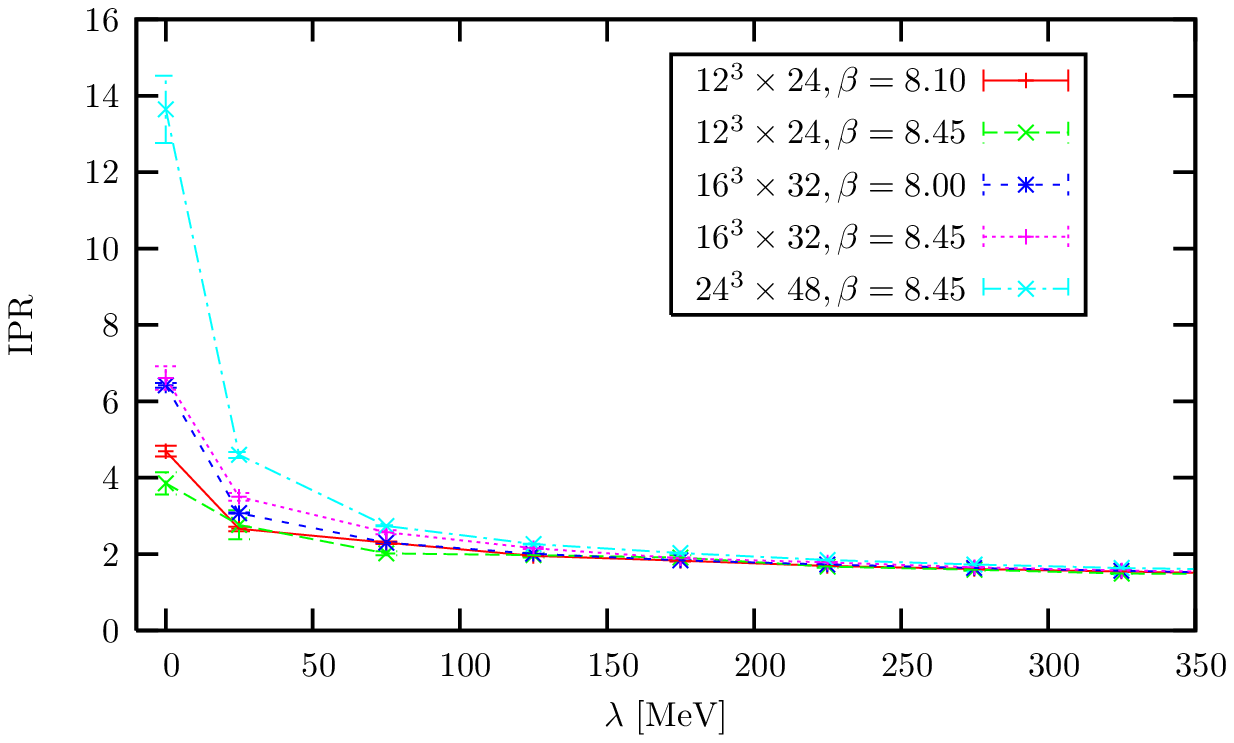}&
\includegraphics[width=7cm]{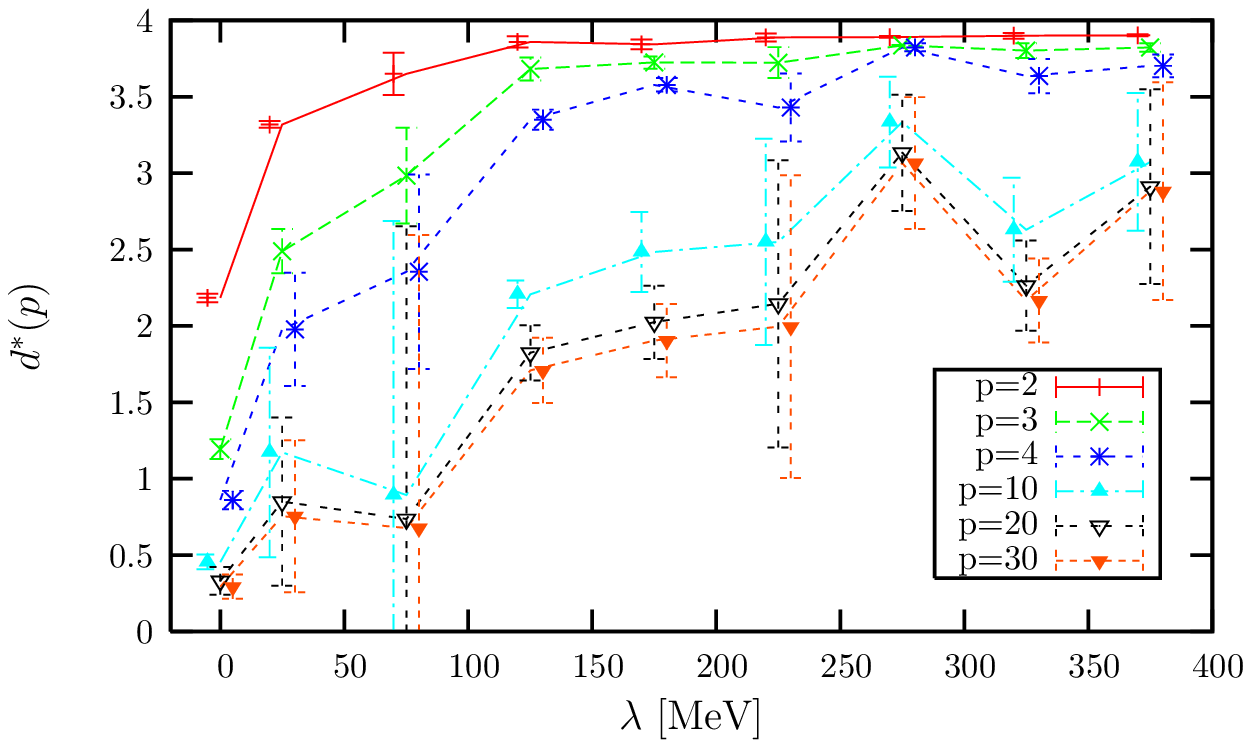}\\
(a) & (b)
\end{tabular}
\end{center}
\caption{(a) The average IPR is shown for zero modes and for 
non-zero modes in $\lambda$ bins of width 50 MeV for the five ensembles.
(b) The fractal dimensionality $d^{*}(p)$ obtained from fits of 
the volume dependence of the averages of the generalized $I_p$ is  
presented for zero modes and for non-zero modes in $\lambda$ bins of 
width 50~MeV for the three ensembles with different volumes and common 
$\beta=8.45$.} 
\label{fig:ipr-2}
\end{figure}

A useful measure to quantify the localization of eigenmodes is the inverse 
participation ratio (IPR), $~I=V \sum_x \rho(x)^2$, with the scalar density 
$\rho(x)=\psi_\lambda^\dagger(x)\psi_\lambda(x)$ for eigenfunctions with
eigenvalues $i\lambda$, which are conventionally normalized as 
$\sum_x \rho(x)=1$. 
In Fig.~\ref{fig:ipr-2} (a) we plot the IPR averaged over bins with a bin width 
$\Delta \lambda = 50$ MeV and for the zero modes  considered separately. 
The average IPR shows an $L$ and $a$ dependence for zero modes and for non-zero 
modes only in the range $\lambda < 150$ MeV. Beyond that the average 
IPR corresponds to 4-dimensionally extended states.
In the theory of the metal-insulator transition, 
a more detailed description of the localization properties at different 
``heights''  of the wave functions has led to consider generalized
IPRs like $I_p= \sum_x \rho(x)^{p}$, averaged over parts of the spectrum 
and different realizations of disorder. 
In any physical application, dimensionalities $d^{*}(p)$ different from the 
embedding dimension $d$ and eventually varying with $p$, can be inferred from 
the volume scaling, 
\begin{center}
\begin{tabular}{lll}
$I_p \propto L^{-d (p-1)}$  & metallic~phase,&extended~wave~functions,\\
$I_p \propto L^{-d^{*}(p) (p-1)}$ & critical~region, &multifractal~wave~functions, \\
$I_p \propto {\rm const}$ &insulator~phase,&localized~wave~functions. 
\end{tabular}
\end{center}
While the dimension for the metallic or insulator phase is $d$ or $0$, respectively, the critical multifractal region in between is characterized by a multitude of fractal dimensions 
$d^{*}(p)$ and also by critical level statistics~\cite{Kravtsov}. 

From a fit of the volume dependence of $I_p$ we find that all eigenstates with 
$\lambda < 200$ MeV in this sense belong to a critical regime. 
From the standard IPR, $I=V~I_2$, we conclude 
that the dimensionality of the zero modes is close to two. In the next bins of 
width 50 MeV each, the dimensionality is near three, in agreement 
with~\cite{MILC} and disagreement with~\cite{Gubarev}~\footnote{
However, in \cite{MILC} the Asqtad Dirac operator on quenched Symanzik improved gauge fields is used, while the analysis in \cite{Gubarev} is done with overlap fermions on quenched SU(2) configurations generated with the standard Wilson action.}.
The IPR of the higher modes is independent of $V$ and $a$, describing the 
observation that the modes freely extend throughout $d^{*}(2) \lesssim d=4$ 
dimensions.

Looking at the higher moments of IPR one can see a decrease of the multifractal 
dimension with increasing $p$ as shown in Fig.~\ref{fig:ipr-2} (b).
This reflects the fact that $I_p$ with $p> 2$ explores regions of higher scalar 
density. For $p=20 \ldots 30$ the emerging fractal dimension seems to converge. 
The envelope reveals that the maxima of the scalar density of zero modes and the 
lowest non-zero modes up to $\lambda = 100$ MeV are characterized by 
$d^{*}(p) < 1$ localization (pointlike peaks), whereas in the  
spectral region  $100~\mathrm{MeV} < \lambda < 400~\mathrm{MeV}$,  
$d^{*}(p)=2$ or $3$ characterizes the maxima of the 
scalar density. In contrast to this, the standard IPR alone would describe 
the modes in the spectral region $\lambda > 150$ MeV as four-dimensional. 
For the zero modes, the results  favor an interpretation in the vortex
picture above, but that needs direct confirmation.
\begin{figure}[ht]
\begin{center}
\begin{tabular}{cccc}
\hspace*{-0.6cm}\includegraphics[width=3.7cm]{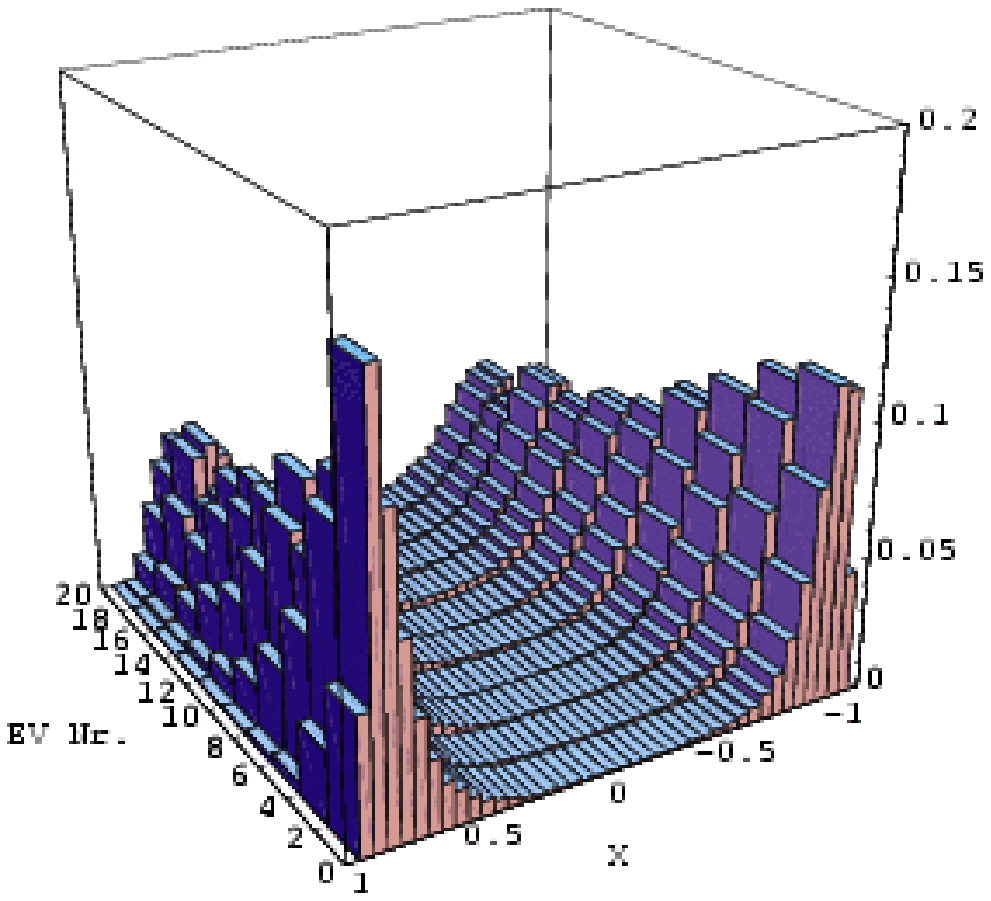}&
\includegraphics[width=3.7cm]{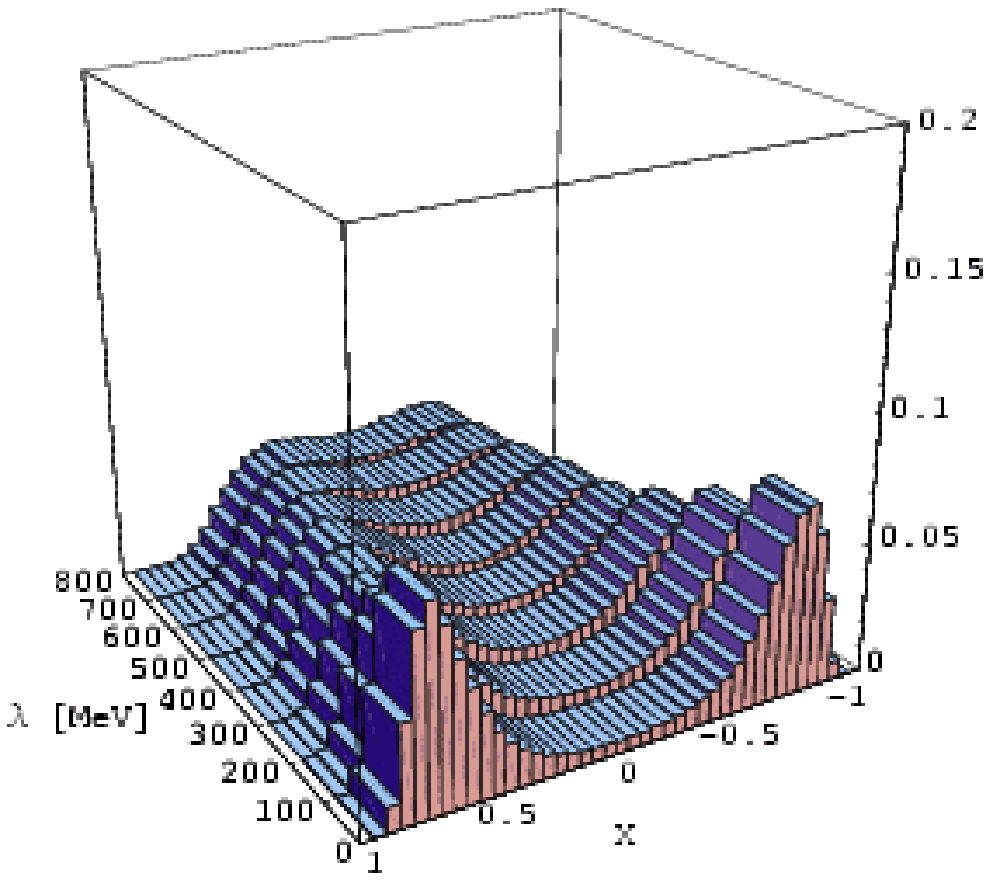}&
\includegraphics[width=3.7cm]{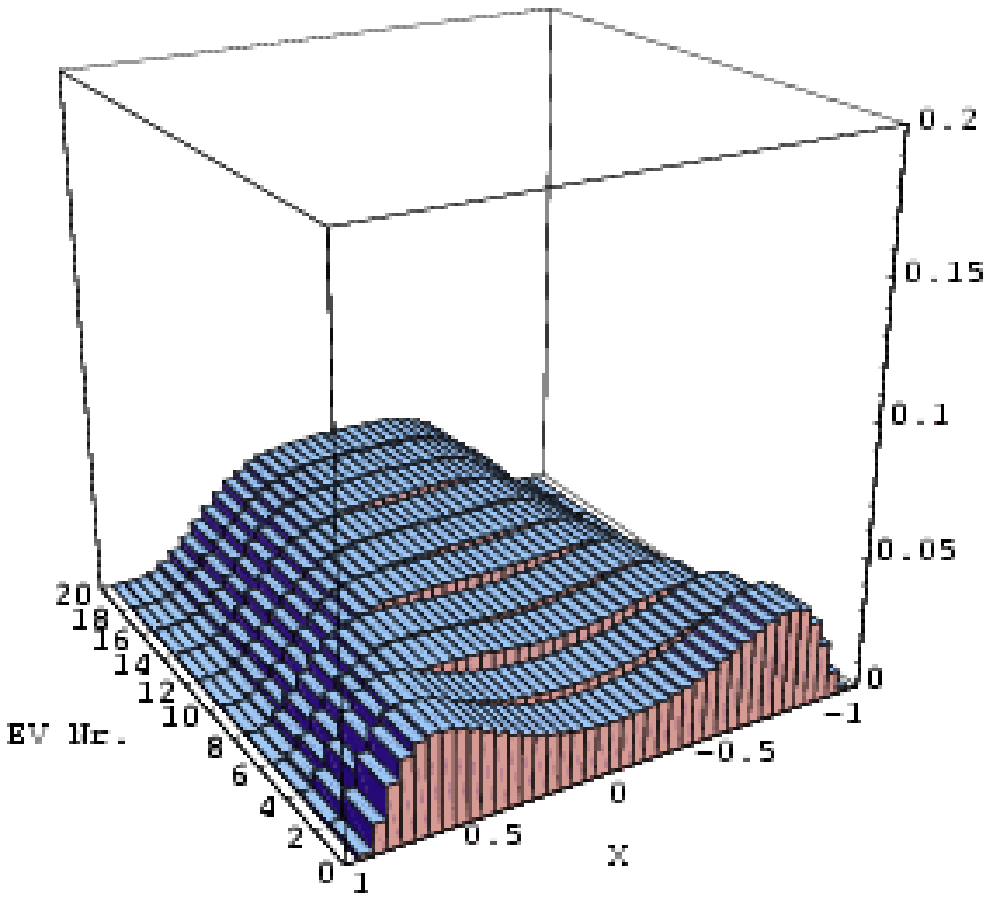}&
\includegraphics[width=3.7cm]{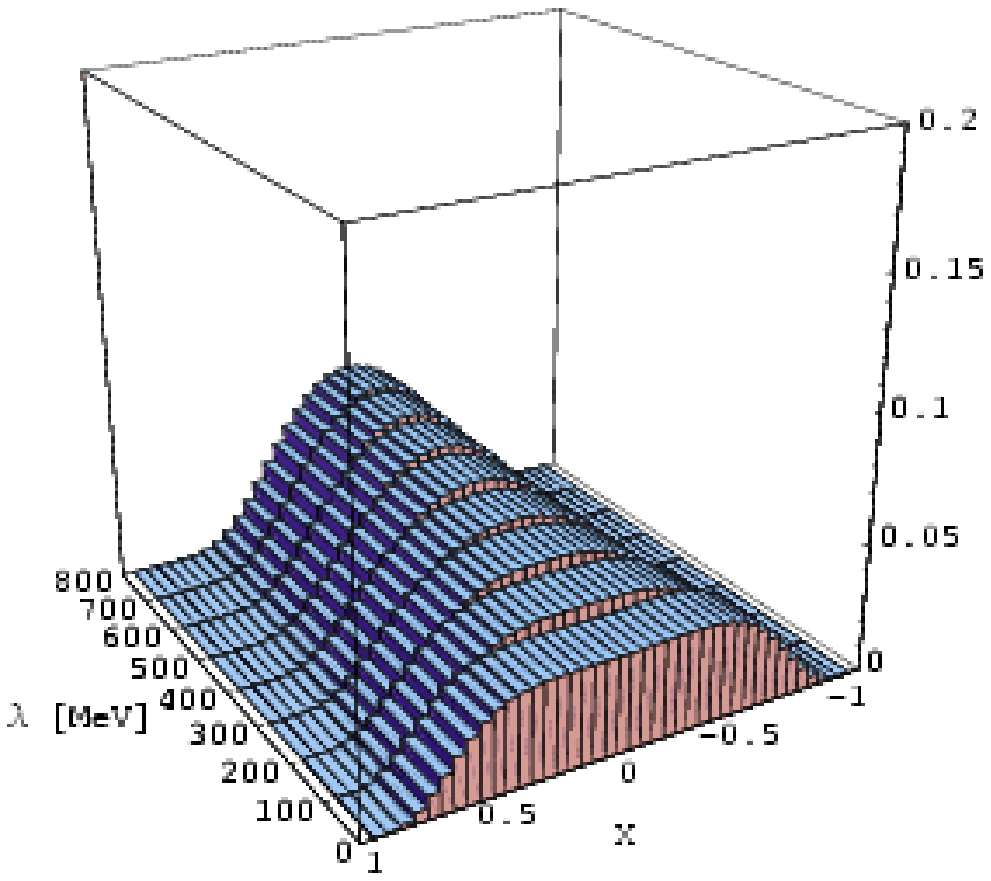}\\
(a) & (b) & (c) & (d) \\
\end{tabular}
\end{center}
\caption{Normalized histograms of the local chirality $X(x)$ in the $Q=0$ subsample 
(consisting of 37 configurations) among the $16^3\times32$ lattices generated 
at $\beta=8.45$.
(a) and (c): for the lowest ten pairs (with positive and negative $\lambda$) averaged over the subsample.
(b) and (d): for all modes from the subsample, averaged over eight $\lambda$ bins of 
width $100$ MeV.
(a) and (b): with a cut for $1/100$ of the sites with biggest scalar density 
$\rho(x)$. (c) and (d): with a cut for $1/2$ of the sites.}
\label{fig:all-mode-loc-chirality-high-dens}
\end{figure}

The local chiral density 
$\rho_{\lambda5}(x) = \psi_{\lambda}^\dagger(x)\gamma_5\psi_{\lambda}(x)$
with $ \sum_x \rho_{\lambda5}(x) = \pm 1$ or $0$ for chiral or non-chiral 
modes, respectively, is equally important.
Whereas the zero modes are entirely chiral, the non-zero modes with globally 
vanishing chirality  may still have a rich local chirality structure 
correlated with the local (anti-)selfduality of the background gauge field. 
A measure proposed by Horvath~\cite{Horvath:2001ir} to describe the local 
chirality $\rho_{\lambda\pm}(x)=\psi_\lambda^\dagger(x)P_{\pm}\psi_\lambda(x)$ 
(with projectors $P_{\pm} = \left( 1 \pm \gamma_5 \right)/2$ onto positive/negative chirality)
is given by 
\begin{equation}
X(x) = \frac{4}{\pi} \arctan \left(r_\lambda(x)\right) - 1 \in [-1,+1] \; ,
\label{eq:arctan}
\end{equation}
with $r_{\lambda(x)}=\rho_{\lambda +}(x)/\rho_{\lambda -}(x)$.
For the chiral zero modes $X(x) \equiv \pm 1$, while for the
non-chiral non-zero modes $X(x)$ is a strongly varying field.

In Fig.~\ref{fig:all-mode-loc-chirality-high-dens} normalized histograms with 
respect to $X(x)$ are shown for the $Q=0$ subsample (37 configurations) of the 
$16^3\times32$ lattices generated at $\beta=8.45$, with cuts including $1/100$ 
and $1/2$ of the sites according to the biggest $\rho(x)$.
For zero modes, the histograms would be concentrated at $X=\pm1$ irrespective
of any cut.
For the non-zero modes, it turns out that with a more restrictive cut in 
$\rho(x)$ the local chirality, {\it i.e.} the concentration of the histogram 
over the included lattice points, can be driven towards $X=\pm 1$.
A similar enhancement can be seen towards the lower modes. For the lowest two 
pairs of non-zero modes all sites with $\rho(x)$ above median show the two 
chiral peaks.   

\section{The UV-filtered gluonic field strength tensor and the topological 
charge density}

The lowest eigenmodes of the overlap operator 
can directly be used as a filter 
of the underlying gauge fields. Following a method proposed by 
Gattringer~\cite{Gattringer}, 
the gluonic field strength tensor can be represented as 
\begin{equation} 
F_{\mu \nu}^a(x) \propto \sum_j \lambda_j^2 \; f^a_{\mu \nu}(x)_j \; .
\label{eq:filter}
\end{equation}
The contribution of the j-th eigenmode 
$f^a_{\mu \nu}(x)_j  
=  -\frac{i}{2} \; \psi_j^\dagger(x) \gamma_\mu \gamma_\nu T^a \psi_j(x)$ 
to the field strength tensor $F_{\mu\nu}^a(x)$ is projected out by means of an
appropriate combination of $\gamma$-matrices with the SU(3) generator $T^a$. 
To study the local (anti-)selfduality of the filtered field strength tensor, 
we analyze the ratio 
\begin{equation}
r(x)=(\tilde s(x)-\tilde q(x))/(\tilde s(x)+\tilde q(x)) \; ,
\end{equation}
with the action density $\tilde s(x) = \Tr\;F_{\mu\nu}(x)~F_{\mu\nu}(x)$  
and the topological density
$\tilde q(x) = \Tr\; F_{\mu\nu}(x)~\tilde{F}_{\mu\nu}(x)$ of the 
filtered field strength summing in
Eq. (\ref{eq:filter}) over different sets of low-lying non-zero modes. 
Using the ratio $r(x)$ we do not need to know the normalization factor
of $\tilde s(x)$ and $\tilde q(x)$. 

\begin{figure}[h]
\begin{center}
\begin{tabular}{cc}
\includegraphics[width=5cm,angle=-90]{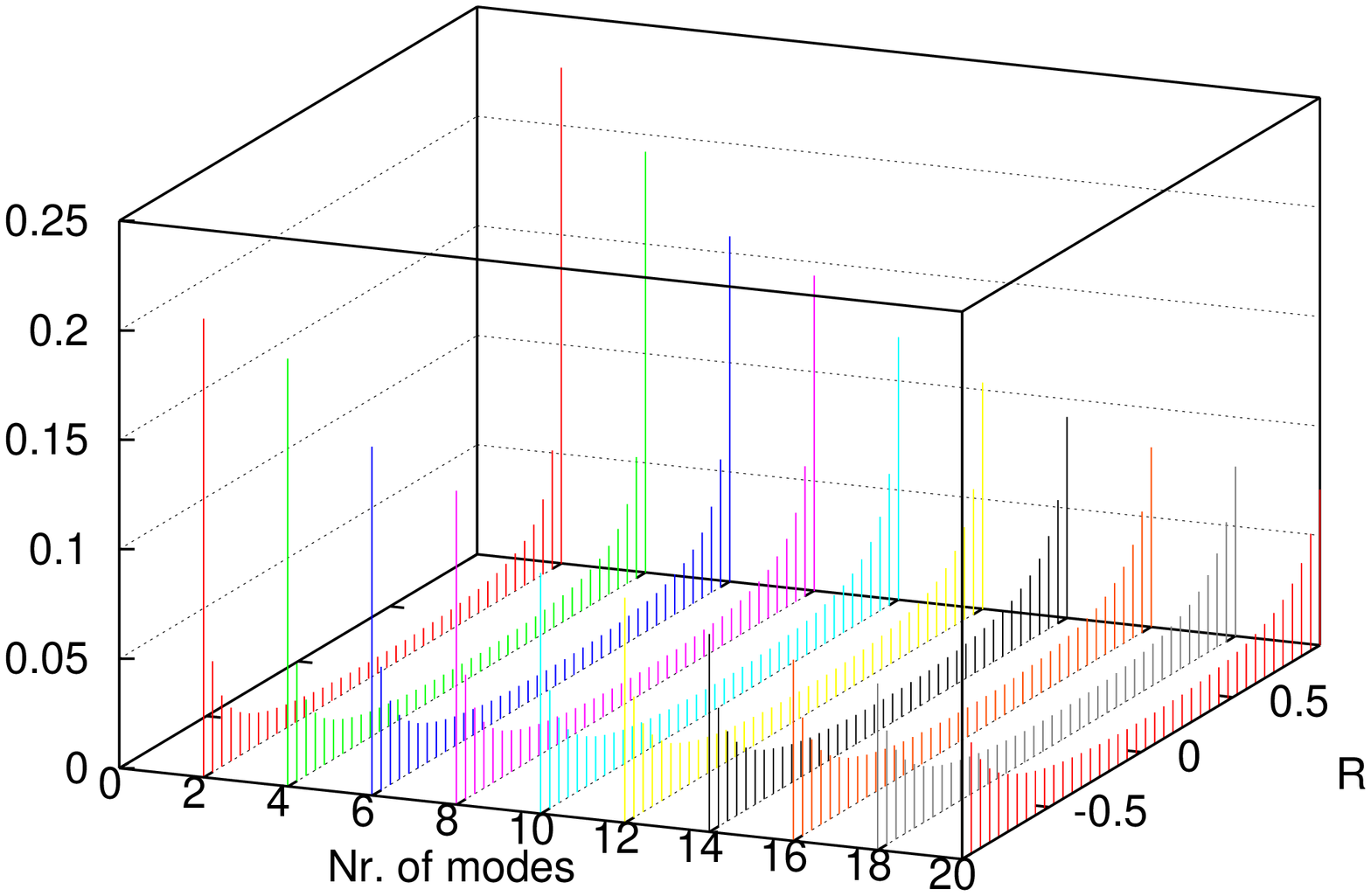}&
\includegraphics[width=5cm,angle=-90]{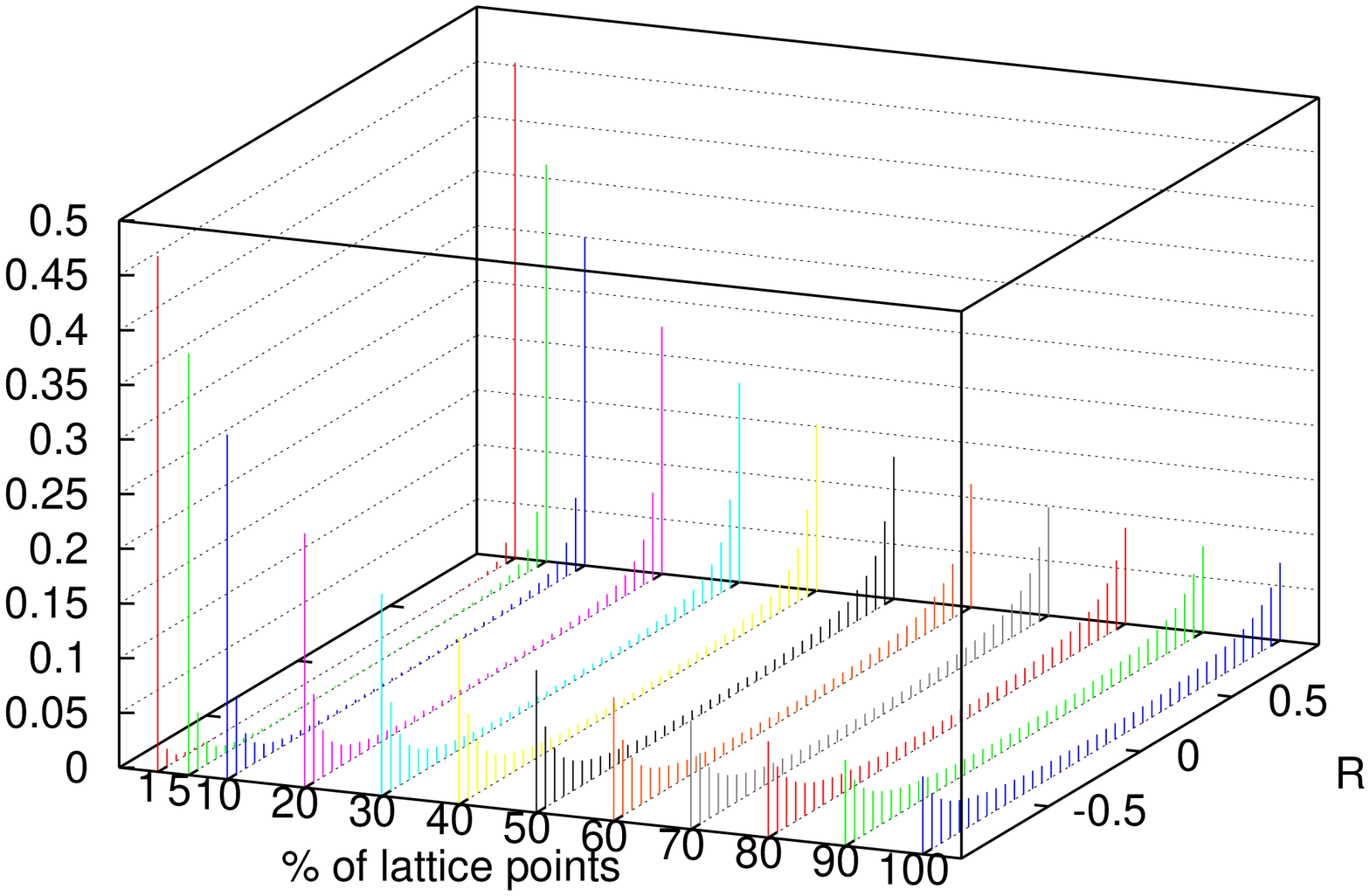}\\
(a) & (b)\\
\end{tabular}
\end{center}
\caption{Normalized histograms with respect to the local (anti-)selfduality $R(x)$ 
of the field strength tensor in the $Q=0$ subsample of 37 configurations 
generated on the $16^3\times32$ lattice at $\beta=8.45$, 
(a) taken over the full volume, but depending on the number of non-zero 
modes (2 - 20) included in the filter, 
(b) applying various cuts according to the action density for a fixed number
(20) of eigenmodes included in the filter.}
\label{fig:Gattringer-histograms}
\end{figure}

Mapping $r(x)$ to $R(x) \in [-1,+1]$ analogously to Eq. (\ref{eq:arctan}), 
one would get $R(x) \equiv \pm 1$ for totally (anti-)selfdual fields. 
In Fig.~\ref{fig:Gattringer-histograms} we present histograms with respect 
to $R(x)$ applying two different cuts with respect to the filtered action 
density $\tilde s(x)$. We observe peaks at $R= \pm 1$ that become weaker 
with the inclusion of more modes. The peaks become more 
pronounced again when one concentrates on lattice points above 
some cut with respect to the filtered action density $\tilde s(x)$.

%\afterpage{\clearpage}

%not used or refered to afterwards ...
%We let $R_{\rm cut} \lesssim 1$ define the isosurfaces $|R(x)|=R_{\rm cut}$
%enclosing domains of almost (anti-)selfdual fields and make a cluster
%analysis with respect to the number, volume and percolation of these domains.

Alternatively to $\tilde q(x)$ based on the filtered field strength tensor 
we also examine the topological charge density as given by the  
formula 
$q(x)=-~\Tr~\left[\gamma_5~\left(1 -\frac{a}{2}~D_N(x,x)\right)\right]$~\cite{Laliena}  
which holds for any $\gamma_5$-Hermitean Dirac operator satisfying the 
Ginsparg-Wilson relation.   
We use it in a two-fold way~\cite{Horvath:2002yn,Horvath:2003yj}.
{\it (i)} We directly calculate the trace of the overlap operator.
The ``all-scale'' density $q(x)$ computed in this numerically very expensive 
way includes charge fluctuations at all scales down to the lattice spacing 
$a$. {\it (ii)} We use the mode-truncated spectral representation \linebreak
$q_{\lambda_{\rm cut}}(x)=-\sum_{|\lambda|<\lambda_{\rm cut}}(1-\frac{\lambda}{2})\; \rho_{\lambda 5}(x)$.

In Ref.~\cite{Ilgenfritz:2005hh} we have begun a cluster analysis of the two forms of 
the topological density. The number of separable clusters of $q(x)$ grows 
without limit with $a \to 0$, and similarly the number of clusters of 
$q_{\lambda_{\rm cut}}$ with increasing $\lambda_{\rm cut}$ at fixed $a$.
We have also seen~\cite{Dublin} that the negativity of the two-point-function 
of the topological charge density~\cite{Horvath:2005cv} (up to a positive core due to the 
non-ultralocality of $q(x)$) is realized for the all-scale density only and 
increases with decreasing lattice spacing $a$. The correlator 
for $q_{\lambda_{\rm cut}}$ describes clusters of the mode-truncated density.
For sufficiently large $\lambda_{\rm cut}$, clusters of different sign are
attractively correlated as demonstrated by a negative tail of the corresponding
correlation function.

\begin{figure}[h]
\begin{center}
\begin{tabular}{ccc}
\hspace*{-0.8cm}
\includegraphics[width=5cm]{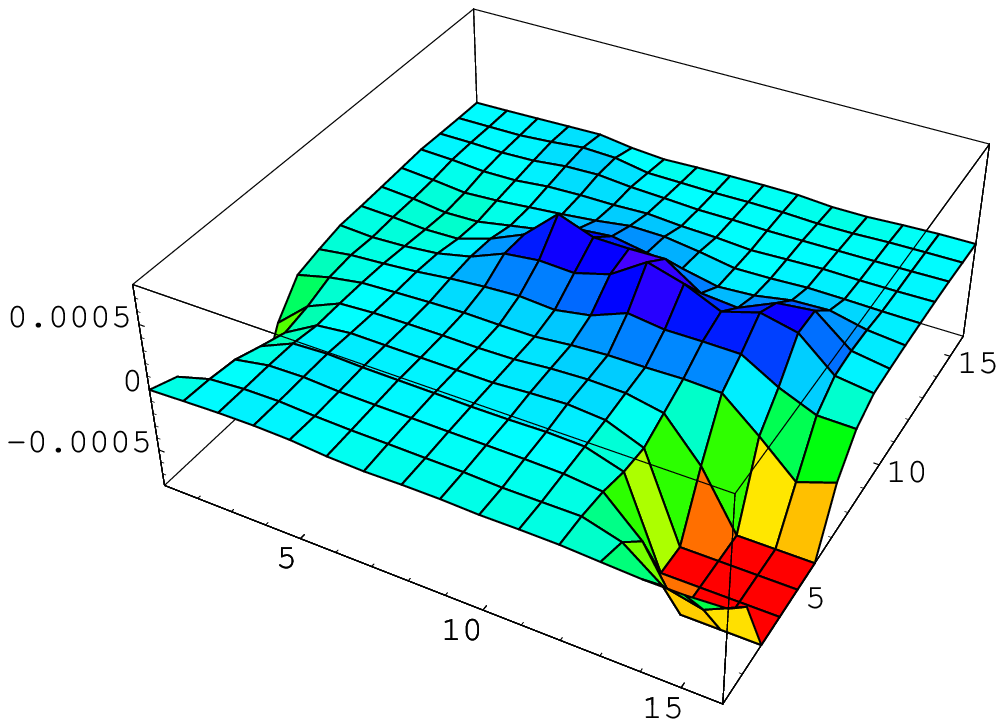}&
\includegraphics[width=5cm]{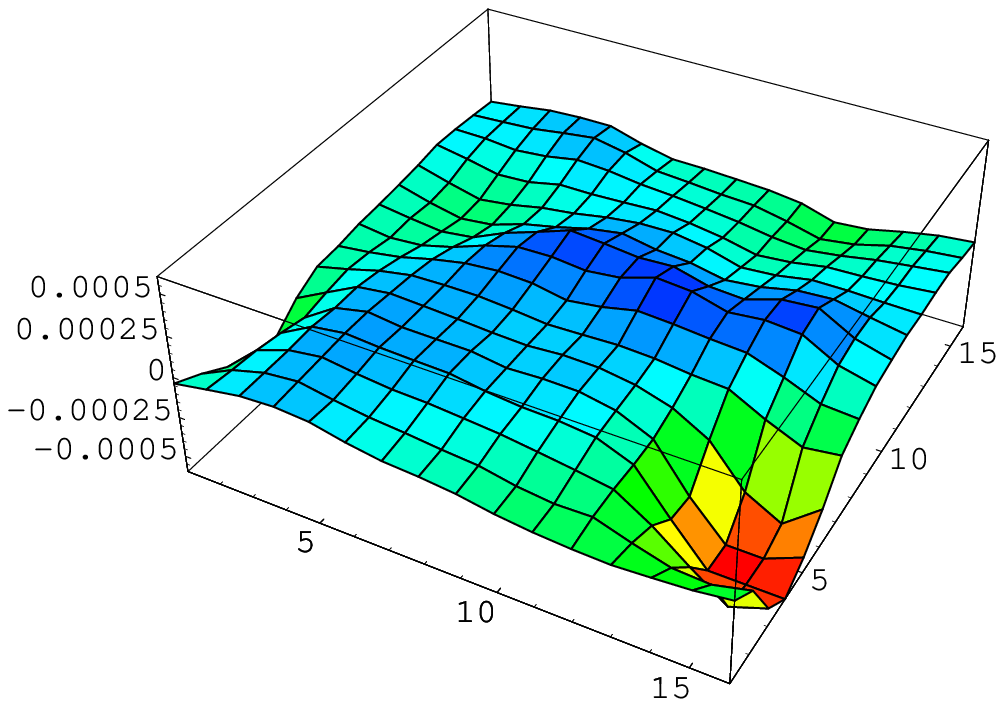}&
\includegraphics[width=5cm]{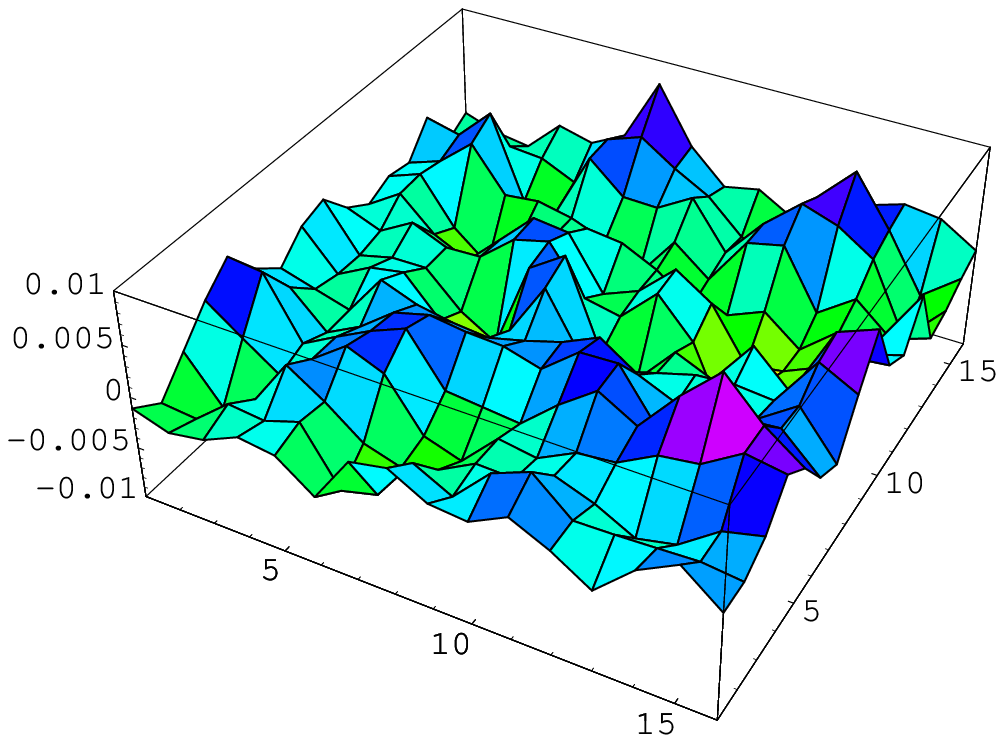}\\
(a) & (b) & (c)\\
\multicolumn{3}{c}{
  \begin{tabular}{cc}
    \includegraphics[width=5cm]{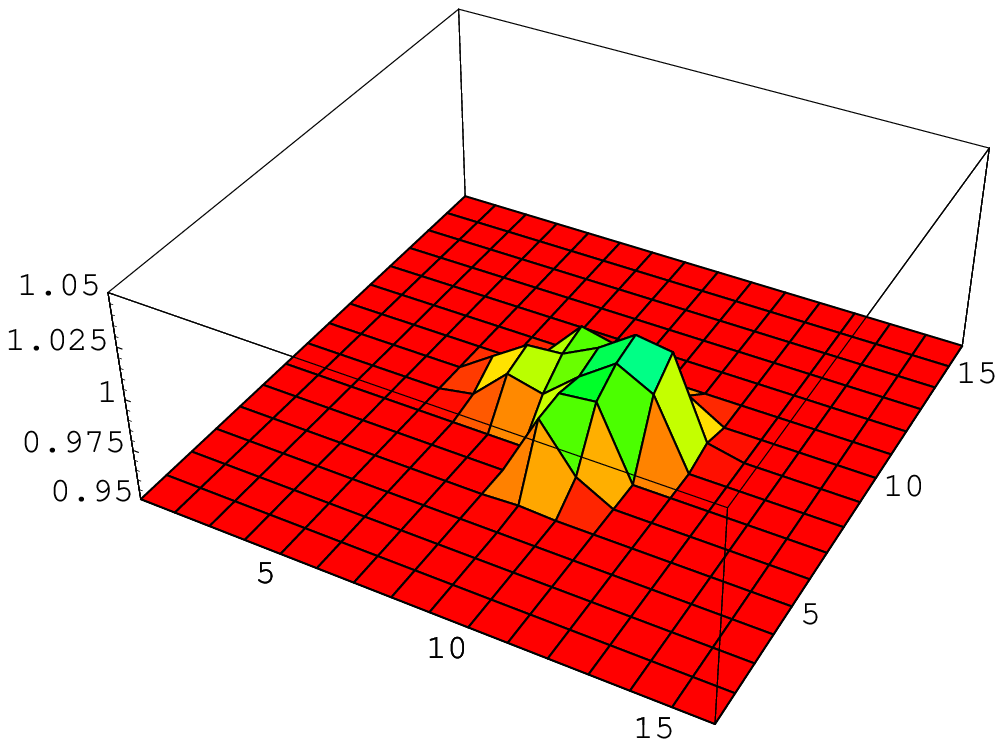}&\includegraphics[width=5cm]{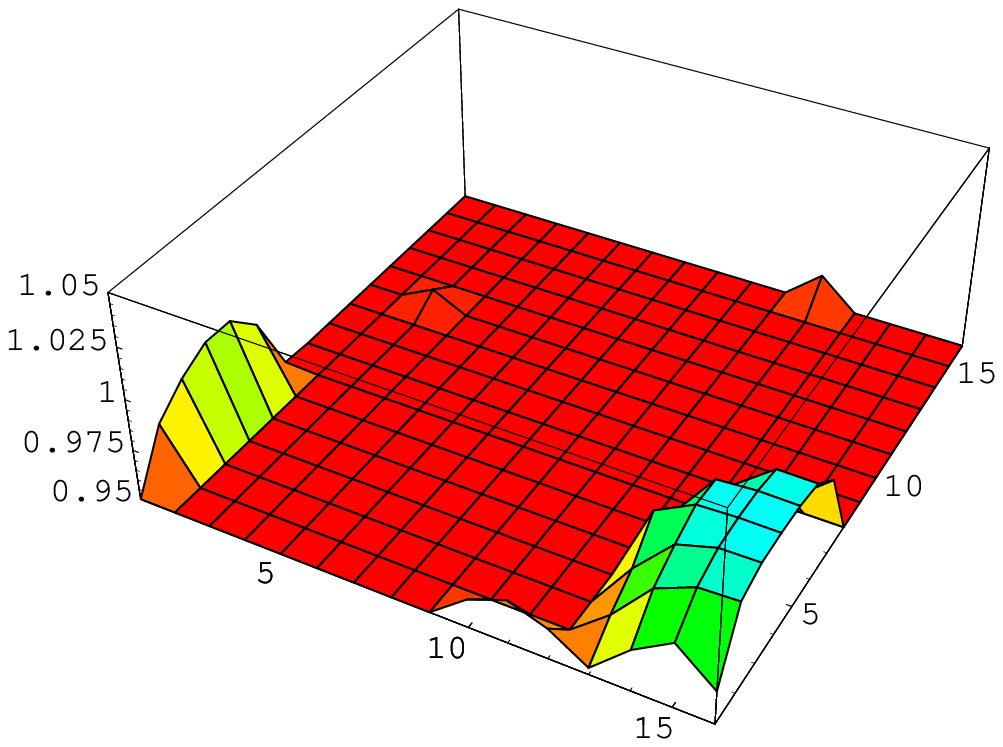}\\
    (d) & (e)\\
  \end{tabular}
}\\

\end{tabular}
\end{center}
\caption{Two-dimensional cuts through a typical configuration
generated at $\beta=8.45$ on a $16^3 \times 32$ lattice.
Direct comparison of (a) $\tilde q(x)$ based on the filtered field strength 
tensor, (b) $q_{\lambda_{\rm cut}}(x)$ as the mode-truncated density and
(c) the all-scale charge density $q(x)$.
Subpanels (d) and (e) show the positive and negative part of the local 
(anti-)selfduality variable $R(x)$. Except for (c) always 18 eigenmodes 
are taken into account.} 
\label{fig:2dcuts}
\end{figure}

In Fig.~\ref{fig:2dcuts} (a)-(c) we compare in a two dimensional cut 
the density $q_{\lambda_{\rm cut}}$ with $\lambda_{\rm cut}=200$ MeV, 
the density $\tilde q(x)$ from the filtered field strength tensor 
and the all-scale density $q(x)$. Viewed is a typical configuration 
generated at $\beta=8.45$ on a $16^3 \times 32$ lattice. Normalizing 
$\tilde q(x)$ in such a way  that 
the maxima of $\tilde q(x)$ and $q_{\lambda_{\rm cut}}(x)$ are equal, 
we see that the two definitions (both including 18 modes) are 
in very good agreement. 
The density of clusters is $O(1 ~\mathrm{fm}^{-4})$, and the number 
of modes necessary to see the clusters approximately corresponds to the 
number of clusters. The topological susceptibility is smaller by a
factor of $1/2$ compared to the dilute instanton-gas approximation. This is an
independent indication for pairing correlations among oppositely charged
clusters of $q_{\lambda_{\rm cut}}$.

In contrast the cut through the all-scale topological density
seems to be completely random. 
It is interesting to notice that structure can be extracted from 
this ``noise'' as well~\cite{Horvath:2005rv,Ilgenfritz:2005hh}.
First of all, cluster percolation at low $q(x)$ leads to the two
thin (thickness of order $O(a)$) sign-coherent global clusters 
for the first time seen by Horvath {\it et al.}~\cite{Horvath:2003yj}. 
At higher $|q(x)|$, 
exploring the interior of the clusters leads to the result that the 
topological landscape is also multifractal, with fractal dimensions 
ranging from $d^{*} = 0.7$ to $d^*=2.6$. 

Finally, Fig.~\ref{fig:2dcuts} (d) and (e) show the positive and 
negative part of the local (anti-)selfduality variable $R(x)$. 
One can see that the peaks of the mode-truncated topological density 
coincide with the domains of (anti-)selfduality. This is evidence 
that with the chosen smearing parameters, extended lumps of high 
topological charge density may be interpreted as (anti-)selfdual objects.
On the other hand, the all-scale density allows one simultaneously
to localize singular fields with fractal dimension $d^{*} < 4$. 
A more quantitative analysis based on our cluster algorithm as drafted 
in \cite{Ilgenfritz:2005hh} will be discussed in detail in a forthcoming 
publication.

\section*{Acknowledgements}

The numerical calculations have been performed on the IBM p690 at HLRN
(Berlin) and NIC (J\"ulich), as well as on the PC farms at DESY Zeuthen, 
LRZ Munich and the University of Munich. 
We thank these institutions for support. Part of this work is 
supported by DFG under contract FOR 465 
(Forschergruppe Gitter-Hadronen Ph\"anomenologie).

\end{document}